# UVB-Radiation-Induced Apoptosis in Jurkat Cells: A Coordinated Fourier Transform Infrared Spectroscopy-Flow Cytometry Study

Deleana Pozzi,[a] Paola Grimaldi,[b] Silvia Gaudenzi,[b] Lucia Di Giambattista,[b] Ida Silvestri,[a] Stefania Morrone[a] and Agostina Congiu Castellano[b,1]

[a] *Dipartimento di Medicina Sperimentale e Patologia and* [b] *Dipartimento di Fisica, Università di Roma "La Sapienza", Roma, Italy*

**Pozzi, D., Grimaldi, P., Gaudenzi, S., Di Giambattista, L., Silvestri, I., Morrone, S. and Congiu Castellano, A. UVB-Radiation-Induced Apoptosis in Jurkat Cells: A Coordinated Fourier Transform Infrared Spectroscopy-Flow Cytometry Study.** *Radiat. Res.* **168,** 698–705 (2007).

We studied the induction of apoptosis in Jurkat cells by UVB radiation (wavelength 290–320 nm) at a dose of 310 mJ/cm$^2$. We combined Fourier transform infrared (FTIR) spectroscopy with flow cytometry to determine whether the combination of both techniques could provide new and improved information about cell modifications. To do this, we looked for correspondences and correlations between spectroscopy and flow cytometry data and found three highly probable spectroscopic markers of apoptosis. The behavior of the wave number shift of both the Amide I β-sheet component and the area of the 1083 cm$^{-1}$ band reproduced, with a high correlation, the behavior of the early apoptotic cell population, while the behavior of the Amide I area showed a high correlation with the early plus late apoptotic cell population.

© 2007 by Radiation Research Society

## INTRODUCTION

In recent years the unique capability of Fourier transform infrared (FTIR) spectroscopy to study molecular cell processes has become more evident. In particular, it represents a nondestructive technique that allows one to obtain simultaneous information on all macromolecules inside a cell population within a few seconds.

Flow cytometry techniques are used to assay biological processes such as cell cycle, apoptosis and necrosis in cells. Recently (*1*), we combined FTIR spectroscopy and flow cytometry to examine apoptosis induced in Jurkat cells by a chemical treatment with actinomycin D. The resulting linear correlation between the flow cytometry data and the behavior of the 2928 cm$^{-1}$ band area in the lipid region, which corresponded to the externalization of the phosphatidylserine, allowed us in that study to consider this band as a spectroscopic marker of processes leading to apoptosis. This mathematical correlation between spectroscopy and flow cytometry data encouraged us to search for other correspondences and correlations that could optimize the measurement of cell modifications. Therefore, we wanted to use the same methodology on the same cell system to investigate the apoptosis induced by other environmental factors such as UVB radiation.

It is known that UVB radiation (wavelength 290–320 nm) has deleterious effects such as skin cancer, suppression of the immune system, and DNA damage by the formation of two lesions, cyclobutane pyrimidine dimers (CPD$_s$) and the 6-4 pyrimidyne-pyrimidone (6-4 PP$_s$) photoproduct at adjacent pyrimidines. These lesions inhibit DNA synthesis and transcription and are generally removed from the genome by mechanisms that include nucleotide excision repair (NER). Cells that have not repaired their DNA are induced to undergo apoptosis as a way to prevent the generation of mutated cell clones (*2–4*).

Even though it is known that UV radiation induces apoptosis in different cell systems, there are no reports in the literature concerning Jurkat cells irradiated with well-defined UVB-radiation doses and wavelengths. In this work we used UVB radiation (wavelength 290–320 nm) at a dose of 310 mJ/cm$^2$.

Apoptosis or programmed cell death is being studied extensively due to its involvement in the development of pathological states. A decreased incidence of apoptosis has been linked with cancer, autoimmune disorders and viral infection, while an increased incidence is an element of many diseases including AIDS, neurodegenerative disorders and ischemic injury. Apoptosis is a regulated process that is controlled by a diversity of internal and extracellular signals that trigger a cascade of complex events and activate different signal transduction pathways. The features accompanying apoptosis, such as the exposure of phosphatidylserine on the outer leaflet of the membrane, membrane blebbing, cytosolic condensation, and DNA nucleosomal fragmentation, are generally dependent on a class of cysteine proteases called caspases. The subsequent formation of well-enclosed apoptotic bodies enables phagocytic cells to remove them, avoiding damage to neighboring cells (*5, 6*).

---

[1] Address for correspondence: Dipartimento di Fisica, Università di Roma "La Sapienza", Piazzale A. Moro 2, 00185 Roma, Italy; e-mail: a.congiu@caspur.it.

## MATERIALS AND METHODS

*Cells and Culture Conditions*

Cells of the lymphoid T-cell line Jurkat, CD3$^+$/CD2$^+$ (*7*), were cultured in RPMI 1640 medium supplemented with 10% fetal bovine serum, 1% penicillin-streptomycin and 1% L-glutamine and were maintained under standard conditions (humidified atmosphere, 95% air, 5% $CO_2$, 37°C). Both the control and irradiated samples were cultured in petri dishes; the unsynchronized cell population has been used for irradiation at a routinely viability of 98% (as measured by Trypan blue exclusion).

*Induction of Apoptosis*

Preliminary cytofluorimetry experiments were carried out to select the optimal UVB-radiation dose and time after UVB irradiation to perform IR measurements.

To induce apoptosis, the cells in uncovered petri dishes were exposed to UVB radiation for different times and at different distances from the source (Philips TL20W/12 lamp emitting 2.1 W at 310 nm). The exposure time of 30 min and a distance of 30 cm produced the maximum percentage of apoptotic cells detected by flow cytometry; the dose, 310 mJ/cm$^2$, was confirmed by radiometer (Gigahertz-Optik GmbH, Germany) measurements. The irradiated cells were then cultured in the incubator for 20, 60, 120, 210, 360 and 420 min until the cytofluorimetry and spectroscopy measurements.

*Flow Cytometry*

Apoptosis was assessed using an Annexin V-FITC apoptosis kit (Bender Medical Systems) and flow cytometry. The percentage of apoptotic cells was determined by green fluorescence emitted by Annexin V-FITC bound to phosphatidylserine exposed to the outer leaflet of the membrane of apoptotic cells. Necrosis was detected using the red fluorescence of propidium iodide (PI) intercalated into DNA (*8*). After extensive washing in PBS, $1 \times 10^6$ cells were rinsed with Hepes buffer, resuspended in Hepes buffer, and incubated at room temperature for 5–15 min in the dark after the addition of Annexin V-FITC; PI was then added to the cell suspension just before the analysis performed using a FACSCalibur (BD Biosciences, San Jose, CA) equipped with an argon-ion laser at an optimal excitation wavelength of 488 nm.

*FTIR Spectroscopy*

The reproducibility of spectra depends on several different parameters (*9*, *10*). Spectra were recorded using an FTIR/410 Jasco spectrometer equipped with a conductive ceramic coil mounted in a water-cooled copper jacket source and a KBr beamsplitter and coupled to an IRT-30 model microscope produced by Jasco Corporation (Japan). The microscope is equipped with a 16× beam condenser and an MCT detector cooled with liquid nitrogen. The bench, microscope and optical path of the instrument are purged continuously with gaseous nitrogen.

The cells for the FTIR measurements were cultured on infrared-transparent CaF$_2$ windows placed in petri dishes and irradiated with UVB light in the same uncovered dishes. The windows had been treated with polylysine to promote cell attachment. After UVB irradiation, treated and control cells were fixed in paraformaldehyde (4% for 20 min), washed in PBS and in distilled water to remove PBS residues, and dried in a desiccator.

*Data Analysis*

All IR spectra were divided into three spectral regions (3000–2800, 1780–1470, 1300–900 cm$^{-1}$), baseline- and smoothing-corrected and area-normalized for every region. The Spectral Manager Analysis software provided by Jasco was used to process the IR spectral data.

For each spectrum, a resolution of 4 cm$^{-1}$ was used, and 256 interferograms were coadded and apodized with a triangular smoothing function before Fourier transformation. The background spectrum was automatically taken into account.

## RESULTS AND DISCUSSION

*Flow Cytometry*

Apoptosis was assessed by flow cytometry using two different fluorescent dyes, FITC-conjugated with Annexin V and PI. Flow cytometry and spectroscopy measurements were made simultaneously at different times after UVB irradiation using cells derived from the same sample. Figure 1 shows the histogram deduced from three different cytometry experiments for the control sample and samples treated with UVB radiation.

The experimental data were analyzed by one-way ANOVA. The results show a significant effect of treatment [$F_{(6,12)} = 5.464$, $P < 0.006$ for live, $F_{(6,12)} = 11.587$, $P < 0.0002$ for early apoptosis, $F_{(6,12)} = 3.070$, $P < 0.047$ for late apoptosis, $F_{(6,12)} = 0.678$, $P < 0.671$ for necrosis).

From the flow cytometry measurements, it is possible to identify four cell populations with different time kinetics: (1) viable cells (white column); the percentage decreases to 36% between 0 and 210 min and increases again to about 53% (low Annexin V-FITC and PI fluorescence); (2) necrotic cells (dark gray column); the percentage (only PI-positive fluorescence cells) is always lower than 4%; (3) cell population with exposure of phosphatidylserine to the membrane outer leaflet and still preserving the plasma membrane integrity (early apoptosis, high Annexin V-FITC and low PI fluorescence, black column); the percentage increases until 210 min, reaching a level of the order of 50%, and then decreases to about 5%; (4) cell population positive both for Annexin V-FITC and for PI fluorescence (late apoptotic cells, the entry of PI into cells, probably due to modifications or damage of the plasma membrane), the percentage exhibits a continuous increase with time (light gray column).

The behavior of live cells in Fig. 1 between 210 and 420 min indicates that at 420 min the population of live cells shows an unexpected increase of about 17%; the significance of this increase was confirmed by the ANOVA results reported above.

*FTIR Spectroscopy*

All FTIR spectra exhibited a good signal-to-noise ratio and were highly reproducible. Figure 2 shows the spectrum of the control sample in two wave-number ranges: from 3000 to 2800 cm$^{-1}$ and from 1780 to 900 cm$^{-1}$. As described in part in our recent paper (*1*) and references therein, the different absorption bands due to the high number of vibrational modes interacting with the physical and chemical environment can be fitted by the addition of the major macromolecule contributions as proteins, lipids, carbohydrates and polynucleotides. The spectral features of each region, attributed to lipids (3000–2800 cm$^{-1}$), proteins

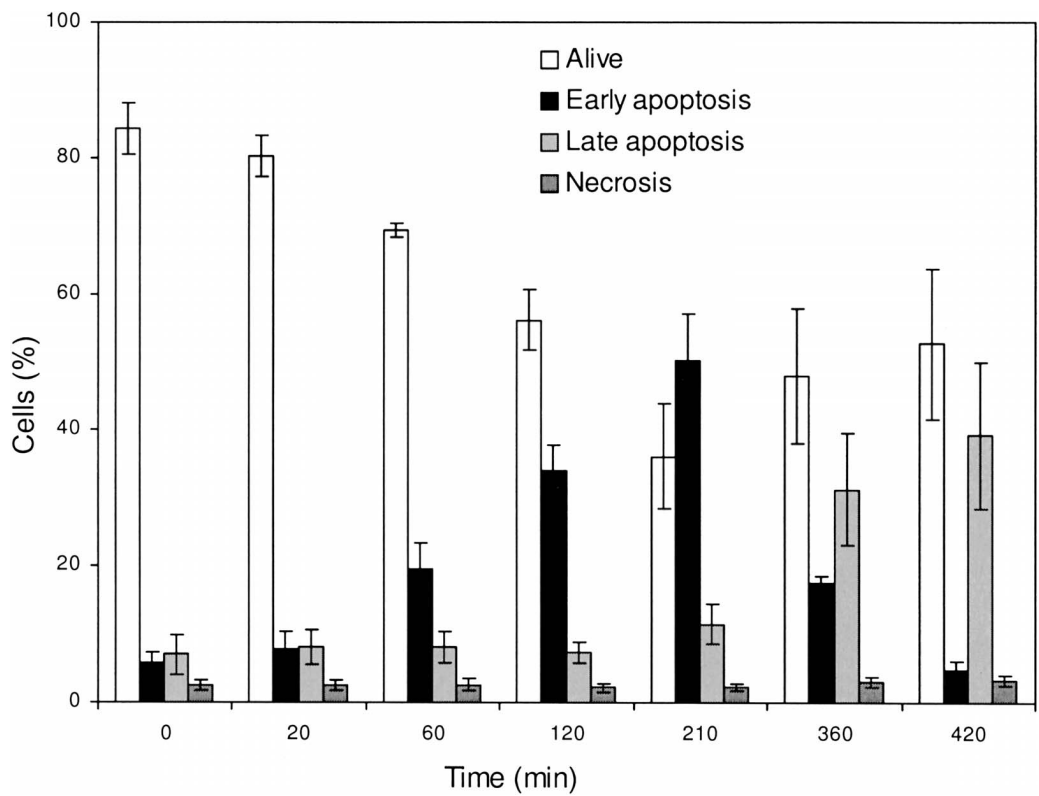

**FIG. 1.** Histogram calculated from three different cytometry experiments for the control and UVB-irradiated samples.

(Amide I and Amide II bands centered at 1660 cm$^{-1}$ and 1541 cm$^{-1}$, respectively), and nucleic acids (1000–1300 cm$^{-1}$ and 950–1000 cm$^{-1}$), were analyzed separately to measure the wave-number shift of the peaks, when this was present, and the total absorption area of the spectral bands.

*1. Lipids*

We analyzed the lipid region (3000–2800 cm$^{-1}$) by comparing the present results with those reported previously (*1*) concerning the induction of apoptosis in Jurkat cells. In our previous work, we observed a time-dependent increase in both the total area of the region and the percentage of apoptotic cells. In particular, the maximum absorbance of the 2928 cm$^{-1}$ band was directly correlated with the percentage of apoptotic cells, which increased progressively until 420 min. In the present experiment, a first analysis of the spectral data immediately revealed that the most significant variations of the wave-number shifts and band area are in the protein and nucleic acid regions, while in the lipid region some of the observed variations lie almost at the limit of the experimental uncertainties. We observed that the 2928 cm$^{-1}$ band is reduced to only one band centered at 2923 cm$^{-1}$ after 120 min. This confirms, with respect to the other bands, an increasing area that is probably correlated with phosphatidylserine externalization.

*2. Proteins*

The strong protein bands, Amide I and Amide II, are centered in the control spectrum at 1660 cm$^{-1}$ and 1541 cm$^{-1}$, respectively. The principal spectral components of the Amide I band are the α-helix structure between 1660 and 1648 cm$^{-1}$, the random-coil structure between 1648 and 1638 cm$^{-1}$, the β-sheet structure between 1640 and 1610 cm$^{-1}$, and the β-turn structure between 1695 and 1660 cm$^{-1}$. In Fig. 3, the Amide I absorption spectra and the second derivatives are shown for the control sample and for the UVB-irradiated samples. The second-derivative analysis of Amide I shows a series of overlapping bands assigned to a mixture of α-helix, β-sheets, β-turns and undefined structures (*11, 12*). The behavior of the α-helix and β-sheet components (respectively at 1660 cm$^{-1}$ and 1627 cm$^{-1}$ in the control sample) is particularly significant because some interesting shifts after UVB irradiation are revealed.

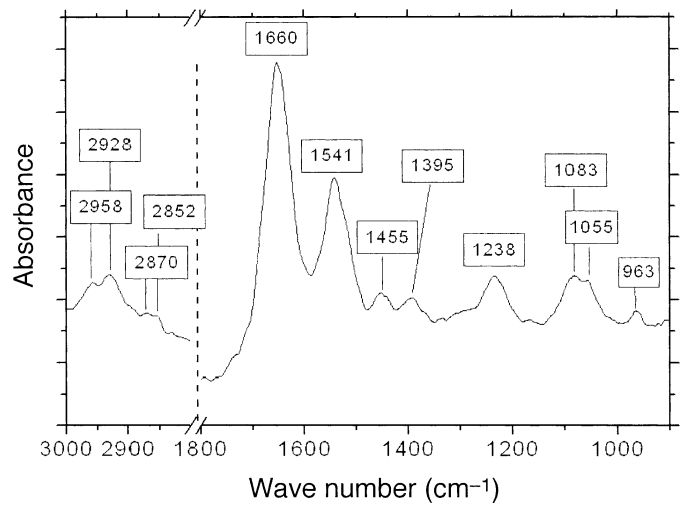

**FIG. 2.** IR spectrum of the control sample. The main features are labeled.

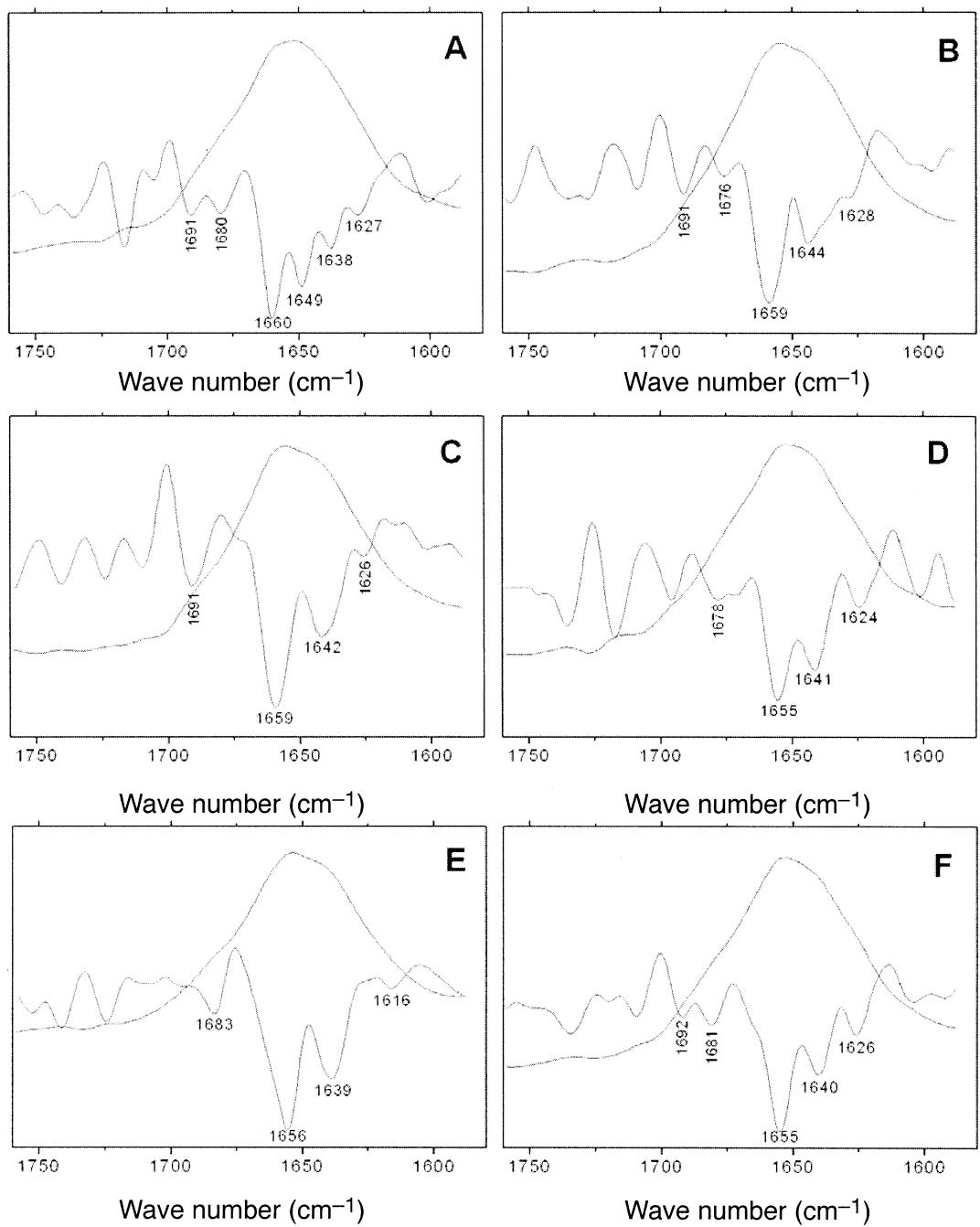

**FIG. 3.** Spectra and second derivatives of the control sample (panel A) and of the samples at 20, 60, 120, 210 and 420 min after UVB irradiation (panels B, C, D, E and F, respectively).

To investigate this behavior further, we studied the wave number as a function of time for both the α-helix and β-sheet structures. While the wave number of the α-helix peak is largely the same as that of the control sample within the experimental uncertainties during the first 60 min, from 120 to 420 min, it has a lower value at about 1655 cm$^{-1}$. Furthermore, the β-sheet peak wave number (Fig. 4) gradually decreases from 1627 cm$^{-1}$ to 1616 cm$^{-1}$ and gradually returns to the control value. In particular, the peak position at 1616 cm$^{-1}$ is present at the same time (210 min) that the highest percentage of apoptotic cells is found by flow cytometry. Therefore, by considering the behavior of early apoptotic cells deduced from Fig. 1, we assumed a correlation between the value of the shift of β-sheet peak wave number and the percentage of apoptotic cells.

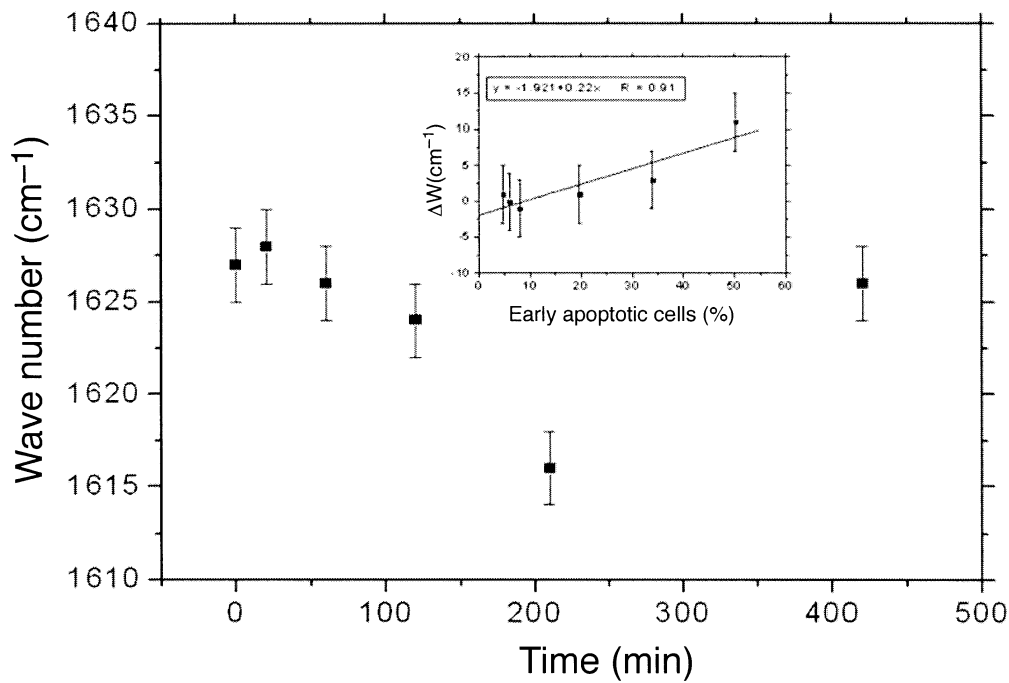

**FIG. 4.** β-sheet peak wave number as a function of time. In the inset, the shift value (Δw: difference between the wave number of the β-sheet peak in the control sample and the wave number of the β-sheet peak in the UVB-irradiated sample) is shown as a function of the percentage of early apoptotic cells.

In the inset of Fig. 4, the shift value (Δw), i.e., the difference between the wave number of the β-sheet peak in the control sample and the wave number of the β-sheet peak in the UVB-irradiated samples, is shown as a function of the percentage of early apoptotic cells. The resulting Pearson's correlation coefficient is 0.91, indicating a probability of about 99% for a linear correlation. This correlation is significant, and we assume that the shift of the β-sheet component is a probable spectroscopic marker of molecular disorder, such as aggregation or denaturation of proteins induced by UVB radiation.

To investigate other possible correlations between the spectroscopy and cytofluorimetry results, we calculated the total area of the Amide I band; the results are shown in Fig. 5, where the uncertainties of the area values have been obtained using the error propagation formula. The comparison of the results shown in Figs. 4 and 5 and the behavior of apoptotic cells (see histogram in Fig. 1) shows the following: (1) The area value increases immediately until 210 min and decreases thereafter. (2) For the shift value of the β-sheet component, the higher area value is present at 210 min in correspondence with the greatest modification displayed in the cytofluorimetry data by UVB-irradiated cells. (3) The kinetics of the spectroscopy and cytofluorimetry responses are similar. This means that the IR spectra are sensitive enough to detect changes in the amounts of vibrating molecular groups showing spectral area modifications when 17% of the early apoptotic cells return to the live cell population. If we correlate the Amide I area with the percentage of apoptotic cells (early plus late), the Pearson's correlation coefficient is 0.84, corresponding to a probability of 96% for linear correlation (inset of Fig. 5); therefore, the Amide I area can be assumed to be another spectroscopic marker that is sensitive to processes that lead to apoptosis.

### 3. Nucleic acids

The comparison of the results for the control sample and UVB-irradiated samples shows that some structures become more evident with time in the irradiated samples, and new bands are observed. The differences with respect to the control spectrum are evident mainly at 60 and 120 min. In particular, 60 min after irradiation, a new band at 983 $cm^{-1}$ (assigned to ribose-phosphate main chain vibration), is observed. This band, at 120 and 210 min, shows a shift to a higher wave number. In addition, the band at 963 $cm^{-1}$ is always present and shows a slight shift to a higher wave number (970 $cm^{-1}$) at 120 min (the second derivatives in the 950–1000 $cm^{-1}$ region are shown in Fig. 6). The 1238 $cm^{-1}$ band assigned to ($\nu_{as}PO_2^-$) stretching, the 1083 $cm^{-1}$ band assigned to ($\nu_s PO_2^-$) stretching, and the 963 $cm^{-1}$ band assigned to C-C/C-O stretching (13) all show peak wave number varying with time. The behavior of the peak position and area is similar for the bands at 1238 $cm^{-1}$ and 963 $cm^{-1}$. These peaks show maximum shifts at 120 min, which is 90 min earlier than the β-sheets. The area of the 1083 $cm^{-1}$ band decreases to a minimum at 120 min (Fig. 7) and then increases back toward the control value, while the percentage of early apoptotic cells becomes about 5% after reaching a maximum at 210 min. At 420 min, the peak positions of the three bands are the same as in the control sample. We looked for a possible correlation between the area of the 1083 cm$^{-1}$ band and the cytofluorimetry results, as shown in the inset of Fig. 7. The correlation coefficient of $-0.77$ indicates a probability of about 93% for a linear correlation.

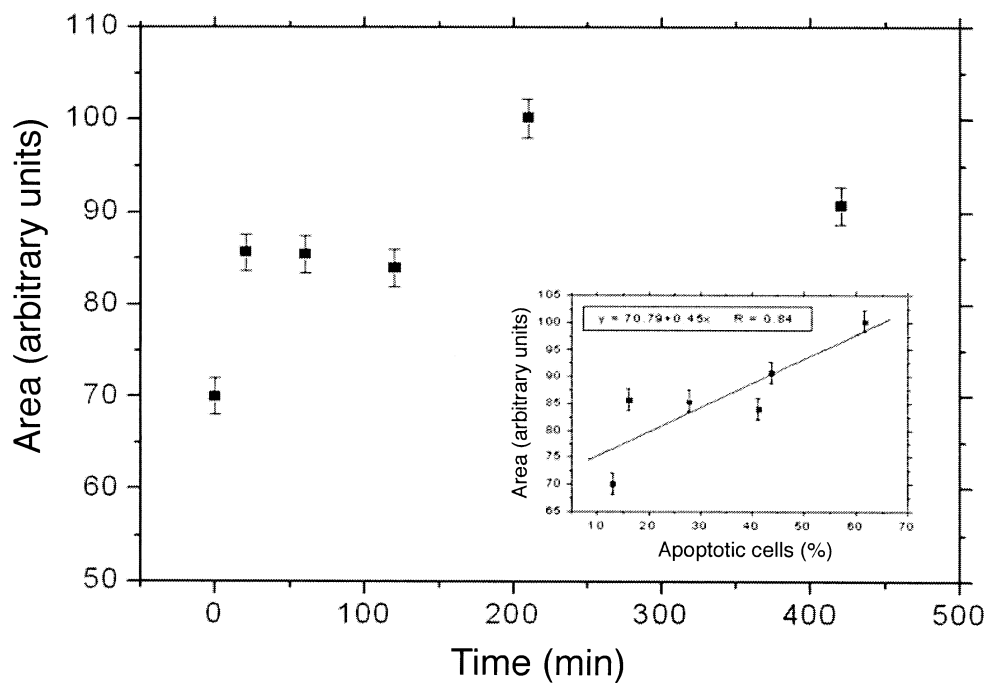

**FIG. 5.** Total area of Amide I band as a function of time. In the inset, the Amide I area as a function of the percentage of early plus late apoptotic cells is shown.

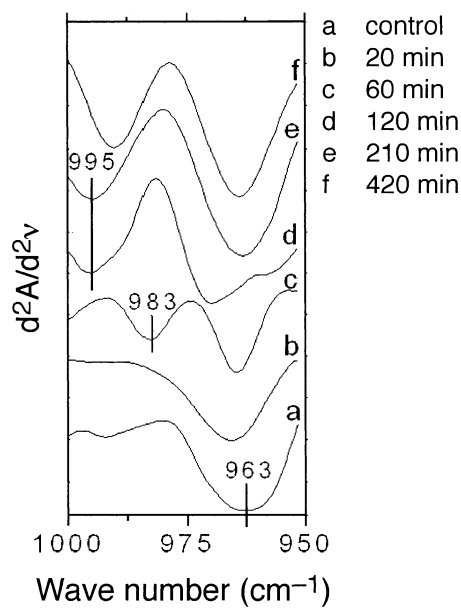

**FIG. 6.** Second derivatives of the spectrum for the 950–1000 cm$^{-1}$ range. Each profile corresponds to the following samples: (a) control sample; (b) 20 min, (c) 60 min, (d) 120 min, (e) 210 min and (f) 420 min after UVB irradiation.

Our experimental results can be summarized as follows:

1. The flow cytometry data show that a UVB-radiation dose of 310 mJ/cm$^2$ induces apoptosis in about 39% of Jurkat cells after 420 min (late apoptotic cells), while the percentage of early apoptotic cells reaches a maximum at 210 min after irradiation before returning to the control level after 420 min. The percentage of live cells decreases until 210 min and then increases by about 17% between 210 and 420 min.

2. Spectroscopy results show that at 120 min after irradiation, the shift toward a lower wave number is maximum for the absorption band at 1660 cm$^{-1}$ assigned to the α-helix. The shift toward a lower wave number is maximum for the absorption band at 1238 cm$^{-1}$ attributed to PO$_2$ stretching. The area of the absorption bands at 1238 cm$^{-1}$ and 1083 cm$^{-1}$ reaches its minimum. At 210 min after irradiation, the shift toward a lower wave number is maximum for the absorption band at 1627 cm$^{-1}$ assigned to β-sheets. The total area of Amide I is maximum. All significant spectroscopic variations appear between 120 and 210 min after irradiation; after 210 min, these values return to those of the nonirradiated sample.

Our experiments showed one significant and two highly probable linear correlations between the spectroscopic and the cytofluorimetric parameters. The shift of the β-sheet peak as a function of the percentage of apoptotic cells, for which correlation factor of 0.91 indicates a probability of about 99% for a linear correlation, can be considered as a spectroscopic marker for monitoring the conformational modifications induced in proteins by a process that leads to apoptosis. The areas of the Amide I and of the 1083 cm$^{-1}$ bands also seem likely to be correlated with the cytofluorimetry results, since the correlation probabilities are 96% and 93%, respectively. We plan to perform other experiments to examine this possibility thoroughly.

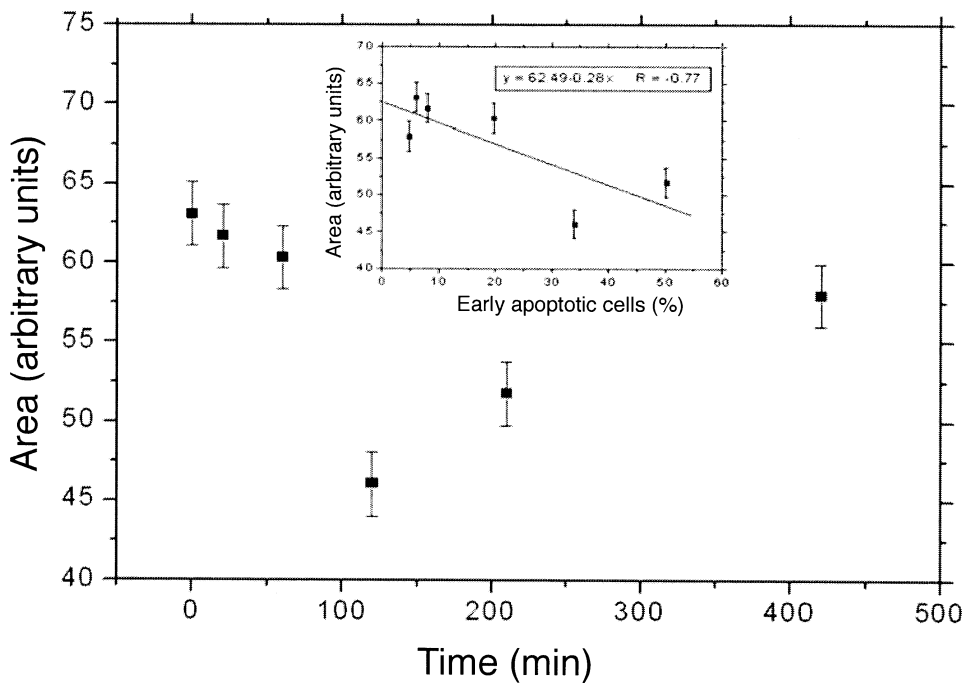

**FIG. 7.** The 1083 cm$^{-1}$ band area as a function of time. In the inset, the 1083 cm$^{-1}$ band area as a function of the percentage of early apoptotic cells is shown.

In the lipid range (3000–2800 cm$^{-1}$), the band assigned to CH asymmetric stretching of CH$_2$ due to two overlapping bands in the control spectrum is reduced to a single band centered at 2923 cm$^{-1}$ after 120 min. This spectral change could be correlated to the major exposure of phosphatidylserine in the outer leaflet of the plasma membrane. The maximum value of the spectral changes is at 120 min, while the maximum value of apoptotic cells is revealed from cytometry data at 210 min; this result is probably related to the different techniques. On one hand, the binding of Annexin V-FITC to the phosphatidylserine molecule is a time-dependent event strictly connected to the externalization of phosphatidylserine; on the other, IR spectroscopy, which is sensitive to aminophospholipid disorders *in toto* from the beginning of the process, appears to be capable of monitoring the conformational changes that occurred at plasma membrane level during events leading to apoptosis in real time in a nonspecific way.

## CONCLUSIONS

The linear correlations we obtained suggest that IR spectroscopy is as sensitive as flow cytometry for monitoring variations in the response of cells to a stimulus that leads to apoptosis. In particular, the behavior of the β-sheets and of the area of 1083 cm$^{-1}$ reproduces with a higher correlation the changes in the percentage of early apoptotic cell populations, while the behavior of the Amide I area shows instead a higher correlation with total apoptotic cells (early + late). Moreover, all the three spectroscopic markers seem sensitive to the presence of the unexpected increase of about 17% in live cells as shown by the recovery of values of the β-sheet shift and 1083 cm$^{-1}$ band area and the partial reversal of the progressive increase in the Amide I area.

Although the data in the literature concerning the UVB radiation dose used are insufficient, we can deduce that a dose of 310 mJ/cm$^2$ in a wavelength range of 290–320 nm in our experimental conditions could be critical to trigger Jurkat cells to undergo apoptosis but does not prevent the live cell population from increasing again to about 17%. This event is probably due to a reversibility mechanism that is interesting not only with respect to the externalization of phosphatidylserine, according to the flow cytometry results, but also with respect to the behavior of proteins and nucleic acids, according to the spectroscopy results.

In summary, our results do not exclude the hypothesis that some events triggered by UVB radiation, in a first phase of that process of apoptosis, could be reversible if the stimulus strength is low and of short duration (*14*). The dependence of the process of apoptosis on the intensity of stimulus has been reported elsewhere (*15*).

This work is the first step of a methodological study. The linear correlations between the spectroscopic and cytofluorimetric parameters need further investigation designed to use them both as "calibration" curves and as a tool for better interpreting experimental results.

## ACKNOWLEDGMENTS

This study was financed by a MIUR 2005 grant. We are grateful to Dott. P. Casale and to Prof. A. Siani of Physics Department, ''La Sapienza'' University and to Dott. M. Borra of the ISPESL Institute for their contributions in the experimental determination of radiation dose and Dr. Marco Costanzi of Neuroscience Institute-CNR for help in statistical analysis. We are grateful to Prof. M. Severini for the supply of the UVB-radiation source and to Dr. Stefano Belardinelli for his assistance in the laboratory experiments.

Received: February 6, 2007; accepted: July 6, 2007

## REFERENCES

1. S. Gaudenzi, D. Pozzi, P. Toro, I. Silvestri, S. Morrone and A. Congiu Castellano, Cell apoptosis specific marker found by Fourier transform infrared spectroscopy. *Spectroscopy* **18**, 415–422 (2004).

2. J. H. Hoeijmakers, Genome maintenance mechanisms for preventing cancer. *Nature* **411**, 366–374 (2001).

3. R. Thyss, V. Virolle, V. Imbert, J. F. Peyron, D. Aberdam and T. Viralle, NF-kB/Egr-1/Gadd 45 are sequentially activated upon UVB irradiation to mediate epidermal cell death. *EMBO J.* **24**, 128–137 (2005).

4. R. Takasawa, H. Nakamura, T. Mori and S. Tanuma, Differential apoptotic pathways in human keratinocyte HaCat cells exposed to UVB and UVC. *Apoptosis* **10**, 1121–1130 (2005).

5. B. Fadeel and S. Orrenius, Apoptosis: a basis biological phenomena with wide-ranging implications in human disease. *J. Int. Med.* **258**, 479–517 (2005).

6. M. Zörnig, A-O. Hueber, W. Baum and G. Evan, Apoptosis regulators and their role in tumorigenesis. *Biochim. Biophys. Acta* **1551**, F1–F37 (2001).

7. U. T. Wirthmueller, T. Kurosaki, M. S. Marakami and J. V. Ravetch, Signal transduction by FcγRIII (CD16) is mediated through the γ chain. *J. Exp. Med.* **175**, 1381–1384 (1992).

8. C. Di Pietro, S. Piro, G. Tabbi, M. Ragusa, V. Di Pietro, V. Zimmitti, F. Cuda, M. Anello, U. Consoli and M. Purrello, Cellular and molecular effects of protons: apoptosis induction and potential implications for cancer therapy. *Apoptosis* **11**, 57–66 (2006).

9. J. R. Mourant, Y. R. Yamada, S. Carpenter, L. R. Dominique and J. P. Freyer, FTIR spectroscopy demonstrates biochemical differences in mammalian cell cultures at different growth stages. *Biophys. J.* **85**, 1938–1947 (2003).

10. K-Z. Liu and H. H. Mantsch, Apoptosis-induced structural changes in leukemia cells identified by IR spectroscopy. *J. Mol. Struct.* **565–566**, 299–304 (2001).

11. T. G. Spiro and B. P. Garber, Laser Raman scattering as a probe of protein structure. *Annu. Rev. Biochem.* **46**, 553–572 (1977).

12. R. W. Williams and A. K. Dunker, Determination of secondary structure of proteins from the amide I band of the laser Raman spectrum. *J. Mol. Biol.* **152**, 783–813 (1981).

13. N. Gault and J. Lefaix, Infrared microspectroscopic characteristics of radiation-induced apoptosis in human lymphocytes. *Radiat. Res.* **160**, 238–250 (2003).

14. C. Dumont, A. Durrbach, N. Bidere, M. Rouleau, G. Kroemer, G. Bernard, F. Hirsch, B. Charpentier, S. A. Susin and A. Senik, Caspase-independent commitment phase to apoptosis in activated blood T lymphocytes: reversibility at low apoptotic insult. *Blood* **96**, 1030–1038 (2000).

15. F. Breuckmann, G. Von Kobyletzki, A. Avermaete, M. Radenhausen, S. Hoxtermann, C. Pieck, P. Schoneborn, T. Gambichler, M. Freitag and P. Altmeyer, Mechanisms of apoptosis: UVA1-induced immediate and UVB-induced delayed apoptosis in human T cells in vitro. *J. Eur. Acad. Dermatol. Venereol.* **17**, 418–429 (2003).